\documentclass[aps,pra,reprint,superscriptaddress,showpacs]{revtex4-2}
\usepackage{amsmath,amsfonts}

\usepackage[colorlinks,linkcolor=blue, urlcolor=blue, anchorcolor=blue, citecolor=blue]{hyperref}
\usepackage[T1]{fontenc}
\usepackage{newtxtext,newtxmath}

\usepackage{graphicx}
\usepackage{bm}

\usepackage{amssymb}
\usepackage{setspace}
\DeclareMathAlphabet{\mathcal}{OMS}{cmsy}{m}{n}  

\begin{document}
	
	
	\title{Spin squeezing generated by the anisotropic central spin model }
	
	\author{Lei Shao}

	\author{Libin Fu}
    \email{lbfu@gscaep.ac.cn}
   	\affiliation{Graduate School of China Academy of Engineering Physics, Beijing 100193, China}



	\begin{abstract}
	Spin squeezing, as a crucial  quantum resource, plays a pivotal role in quantum metrology, enabling us to achieve high-precision parameter estimation schemes. Here we investigate the spin squeezing and the quantum phase transition in an anisotropic central spin system. We find that this kind of central spin systems can be mapped to the anisotropic Lipkin-Meshkov-Glick model in the limit where the ratio of transition frequencies between the central spin and the spin bath tends towards infinity. This property can induce a one-axis twisting interaction and provides a new possibility for  generating spin squeezing. We separately consider generating spin-squeezed states via the ground state and the dynamic evolution of the central spin model. The results show that the spin squeezing parameter improves as the anisotropy parameter decreases, and its value scales with system size as $N^{-2/3}$. Furthermore, we obtain the critical exponent of the quantum Fisher information around the critical point by numerical simulation, and find  its value tends to $4/3$ as the frequency ratio and the system size approach infinity. 
 	This work offers  a promising scheme for generating spin-squeezed state and paves the way for potential advancements in  quantum sensing.
	

	\end{abstract}
	
	
	\maketitle
	
	\section{Introduction}
	
	As the understanding of the quantum world continues to deepen, quantum technology is ushering in  in a series of revolutionary changes in the real world. Notably, quantum precision measurement is progressively gaining prominence, especially playing a crucial role in critical areas such as 
	 biophysics~\cite{TAYLOR20161,Taylor2013,NASR20091154,Morris2015}, inertial sensors~\cite{Qvarfort2018,doi:10.1126/sciadv.add3854}, and measurement of physics constants~\cite{PhysRevLett.100.050801,PhysRevLett.103.050402,PhysRevA.65.033608}. In the realm of quantum metrology, a core goal is to improve the measurement precision of parameters of interest. As a result, a central focus of research lies in exploring how to utilize the nonclassical properties of quantum resources to achieve heightened precision. In the past decades, quantum spin squeezing has garnered considerable attention as an important concept in the field of quantum information, particularly within the domains of quantum metrology and quantum sensing. Due to its capability to reduce quantum fluctuations of a specific spin component, spin squeezing can enhance the precision of measurements, thus  spin-squeezed states are widely applied to quantum precision measurement, such as atomic clock~\cite{PhysRevA.71.043807,PhysRevLett.83.2274,PhysRevA.81.043633,PhysRevLett.92.230801}, Ramsey spectroscopy~\cite{PhysRevA.53.467,10.1007/3-540-45338-5_4,PhysRevA.59.R922,PhysRevLett.86.5870,PhysRevLett.93.163602} and gravitational wave interferometers~\cite{WALLS1981118,PhysRevA.70.033601,Goda2008}.  Furthermore, investigating the spin squeezing is also helpful  to gain a deeper understanding of the correlations, entanglement, and  quantum information processing among particles in the quantum world~\cite{PhysRevA.68.012101,PhysRevLett.86.4431,PhysRevA.79.042334,Estève2008}.
	
	The generation of spin squeezing involves two categories~\cite{MA201189}, including nonlinear atomic-atomic collisions in Bose-Einstein condensates (BECs)~\cite{Sørensen2001,Orzel2001,PhysRevA.63.055601,PhysRevA.66.043621,PhysRevA.64.013616,PhysRevA.65.033625,PhysRevA.67.013607,PhysRevA.70.063612,PhysRevA.72.033612,PhysRevA.71.043603,PhysRevLett.99.170405,PhysRevA.76.043621} and atomic-photon interactions~\cite{PhysRevA.46.R6797,PhysRevLett.83.1319,Julsgaard2001,RevModPhys.82.1041,PhysRevA.39.1962,PhysRevA.54.5327,PhysRevA.60.2346,PhysRevLett.80.3487,PhysRevA.62.063812,PhysRevA.69.043810,PhysRevA.81.021804,PhysRevA.104.053517,doi:10.1126/science.1095374,PhysRevLett.104.073604,PhysRevA.62.033809,PhysRevLett.85.1594,PhysRevLett.104.073602}. The former exploits the nonlinear effects of atomic collisions to generate spin-squeezed states, while the latter
	mainly involves utilizing squeezing transfer from the optical field to spin ensemble~\cite{PhysRevLett.83.1319,PhysRevA.62.033809,PhysRevA.69.043810,PhysRevLett.85.1594,PhysRevA.104.053517}, the coherent feedback of the optical cavity~\cite{PhysRevA.81.021804,PhysRevLett.104.073602} or  quantum nondemolition (QND) measurement of the output light~\cite{PhysRevA.60.2346,PhysRevLett.85.1594,doi:10.1126/science.1095374,PhysRevLett.104.073604}.
	
	 Utilizing bosonic atoms to generate spin squeezing can result in more losses due to collisions~\cite{PhysRevLett.129.090403,RevModPhys.90.035005,Kolkowitz2017}, hence it is essential to explore the use of spinful fermions to achieve spin squeezing. Due to the similarity between the light-matter interaction and the spin-spin interaction, the central spin systems can also induce one-axis twisting (OAT) interaction under certain conditions, thus it can be regarded as a promising alternative model to realize spin squeezing. The scheme proposed in \cite{PhysRevLett.107.206806} also utilizes a similar approach to achieve spin squeezing in a quantum dot composed of an electron and a large ensemble of nuclear spins. However, the mechanism of inducing the effective one-axis twisting  interaction in central spin systems is still ambiguous, thus, we employ the Schrieffer-Wolff transformation to rigorously illustrate the generation mechanism of OAT
	interaction and extend it to the anisotropic case.  We find that when the ratio of the transition frequency of the central spin to that of the bath spin tends to infinity, the anisotropic central spin model can be mapped to the anisotropic Lipkin-Meshkov-Glick (LMG) model~\cite{LIPKIN1965188}, where the OAT interaction in the Hamiltonian can dynamically generate spin squeezing . In addition, there exists a quantum phase transition (QPT) in this anisotropic central spin model, and its quantum criticality can be viewed as a quantum resource for quantum metrology. Thus we investigate the finite-size behavior around the critical point and numerically calculate the critical exponent of the quantum Fisher information (QFI). It is found that when the frequency ratio tends to infinity, the critical exponents of the central spin model become equal to those of the LMG model, however, as the frequency ratio decreases, the critical exponent will decay to zero.
	\setlength{\parskip}{1em}
	
	This article is organized as follows. In Sec.~\ref{sec:levelI}, we give the derivation of the mapping between the anisotropic central spin model and the anisotropic LMG model. In Sec.~\ref{sec:levelII}, we investigate the spin squeezing and quantum phase transitions of the central spin model under both isotropic and anisotropic conditions. In Sec.~\ref{sec:levelIII}, we discuss the quantum Fisher information and its critical exponent for the ground state in the anisotropic central spin model. Finally, we give a conclusion in Sec.\ref{sec:levelIIII}.

	\section{\label{sec:levelI}MODEL}
	The central spin model can be described as an ensemble of $N$ identical spin-$\frac{1}{2}$ particles with a total spin of $I=N/2$ interact with a central spin-$\frac{1}{2}$ particle. This model is discussed in the Ref.~\cite{PhysRevA.87.052323,PhysRevB.99.174308,PhysRevA.107.013714,PhysRevLett.88.186802,PhysRevLett.88.186802,PhysRevB.82.161308,PhysRevB.76.014304,PhysRevLett.118.150504} and its Hamiltonian is written as (we set $\hbar=1$)
    \begin{equation}\label{S2-1}
	\begin{split}
	H=&\frac{\Omega}{2}\sigma^{(0)}_{z}+\frac{\omega}{2} \sum_{k=1}^{N}\sigma_{z}^{(k)}+
	\sum_{k=1}^{N}\left(\frac{A_{x}}{2} \sigma^{\left(k\right)}_{x}\sigma^{\left(0\right)}_{x}\right.\\
	&\left.+\frac{A_{y}}{2}\sigma^{(k)}_{y}\sigma^{\left(0\right)}_{y}+\frac{A_{z}}{2}\sigma^{\left(k\right)}_{z}\sigma^{\left(0\right)}_{z}\right),
	\end{split}
	\end{equation} 
	where $\Omega$ and $\omega$ are the transition frequencies  of the central spin and bath spins induced by  an external magnetic field, respectively. And $A_{i}$ ($i=x,y,z$) represents the strength of interaction between the central spin and bath spins in different directions. Here $\sigma^{\left(0\right)}_{i}$ represents the Pauli operator of the central spin and $\sigma^{(k)}_{i}$ ($k\neq0, i=x,y,z$) is the $i$th Pauli operator of the bath spins. Central spin model is widely used to study the spin-spin interactions in quantum dots~\cite{PhysRevLett.88.186802,PhysRevLett.88.186802,PhysRevB.82.161308,PhysRevB.76.014304} and nitrogen-vacancy center~\cite{PhysRevLett.118.150504}. Some intriguing physical phenomena, such as collapse and revival~\cite{PhysRevA.87.052323,PhysRevB.99.174308}, superradiance effect~\cite{PhysRevA.87.052323}, and dissipative phase transitions~\cite{PhysRevA.86.012116}, emerge within this model. In this article, we mainly focus on the anisotropic  model, that is, the coupling strength of the $X$ and $Y$ directions is different. Here we set  $A_{x}=\left(1+\lambda\right)A$, $A_{y}=\left(1-\lambda\right)A$, and $A_{z}=0$, and Eq.~\eqref{S2-1} can be rewritten as  
	\begin{equation}\label{S2-2}
	H=\frac{\Omega}{2} \sigma_{z}+\omega I_{z}+A\left[\left(I_{+}\sigma_{-}+I_{-}\sigma_{+}\right)+\lambda\left(I_{+}\sigma_{+}+I_{-}\sigma_{-}\right)\right],
	\end{equation}
	where $\sigma_{z}^{\left(0\right)}\equiv\sigma_{z}$,  $I_{z}=\frac{1}{2}\sum_{k=1}^{N}\sigma_{z}^{(k)}$, and $\lambda$ is the anisotropy parameter. Note that the Hamiltonian in Eq.~\eqref{S2-2} commutes with the parity operator $\Pi=\exp\left[i\pi\left(\sigma_{+}\sigma_{-}+I_{z}+N/2\right)\right]$, i.e. $\left[\Pi,H\right]=0$, indicating that it possesses the $Z_{2}$ symmetry. However, if $\lambda\neq1$  then it can not be solved with a closed subspace,  which complicates further analysis of this model. To this end, we  apply a Schrieffer-Wolff transformation $e^{S}$ with $S=-i\left(1+\lambda\right)/\Omega I_{x}\sigma_{y}+i\left(1-\lambda\right)/\Omega I_{y}\sigma_{x}$ to Eq.~\eqref{S2-2}, and in the limit of $\eta=\Omega/\omega\rightarrow\infty$ we obtain 
	\begin{equation}\label{S2-3}
	H=\frac{\Omega}{2}\sigma_{z}+\gamma_{z} I_{z}+\frac{1}{N}\left(\gamma_{x}I_{x}^{2}+\gamma_{y}I_{y}^{2}\right)\sigma_{z},	
	\end{equation}
	where  
	\begin{align}
    \gamma_{x}&=\frac{\tilde{g}^{2}\omega\left(1+\lambda\right)^{2}}{4},\ \gamma_{y}=\frac{\tilde{g}^{2}\omega\left(1-\lambda\right)^{2}}{4},\label{S2-4}\\ 
    \gamma_{z}&=\omega-\widetilde{g}^{2}\frac{\left(1+\lambda\right)\left(1-\lambda\right)\omega}{4N},\label{S2-5}\\
    \tilde{g}&=\frac{2A\sqrt{N}}{\sqrt{\Omega\omega}}\label{S2-6}.
	\end{align}
	Here $\tilde{g}$ in Eq.~\eqref{S2-6} is a dimensionless coupling
	strength parameter. It can be seen that the Hamiltonian in Eq.~\eqref{S2-3} is block-diagonal with respect to spin states and its low-energy effective Hamiltonian in the spin down subspace is
	\begin{equation}\label{S2-7}
	H_{\downarrow}=-\frac{\Omega}{2}+\gamma_{z} I_{z}-\frac{1}{N}\left(\gamma_{x}I_{x}^{2}+\gamma_{y}I_{y}^{2}\right).
	\end{equation} 
	 One can see that Eq.~\eqref{S2-7} is exactly the Hamiltonian of the anisotropic LMG model. In other words, there exists a mapping between the anisotropic  model  and  the  LMG model in the limit of $\eta\rightarrow\infty$. The detailed derivation of the above process is presented in Appendix \ref{Appendix:A}. A Similar method for generating spin squeezing has been discussed in Ref.~\cite{PhysRevLett.107.206806}, where they provided an approximate relationship. In contrast, we present a more detailed mapping relationship and derivation process here.
	 
	\section{\label{sec:levelII}SPIN SQUEEZING AND QUATNUM PHASE TRANSITIONS IN CENTRAL SPIN MODEL }
	
	In this section, we will discuss the spin squeezing and quantum phase transtion in the anisotropic central spin model. We will begin by introducing two commonly used definitions of spin squeezing proposed by Kitagawa \emph{et al}.~\cite{PhysRevA.47.5138} and Wineland \emph{et al}.~\cite{PhysRevA.50.67}, which are given by 
	\begin{equation}\label{S2-8}
	\xi_{S}^{2}=\frac{4{\rm min}\left(\Delta^{2} I_{\vec{n}_{\perp}}\right)}{N},	
	\end{equation}  
	and
	\begin{equation}\label{S2-9}
	\xi_{R}^{2}=\frac{N{\rm min}\left(\Delta^{2} I_{\vec{n}_{\perp}}\right)}{\left|\left\langle \vec{I}\right\rangle \right|^{2}},
	\end{equation}  
	where $\vec{n}_{\perp}$ is the direction perpendicular to the mean spin direction $\left\langle\vec{I}\right\rangle$ and $\Delta^{2} I_{\vec{n}_{\perp}}$ is the variance of  $I_{\vec{n}_{\perp}}=\vec{I}\cdot\vec{n}_{\perp}$. And if $\xi_{S}^{2}<1$ or $\xi_{R}^{2}<1$, it implies that this state is a spin-squeezed state. In the following subsections, we will analyze the spin squeezing in the isotropic ($\lambda=0$) and anisotropic ($\lambda\neq0$) cases separately.
	\subsection{\label{subsec:levelA}Isotropic case}
	
	For $\lambda=0$, the Hamiltonian in Eq.~\eqref{S2-2} becomes
	\begin{equation}\label{S2-10}
		H=\frac{\Omega}{2} \sigma_{z}+\omega I_{z}+\frac{\tilde{g}}{2}\sqrt{\frac{\Omega\omega}{N}}\left(I_{+}\sigma_{-}+I_{-}\sigma_{+}\right).
	\end{equation}
	We can find that the above Hamiltonian remains invariant under the action of  $\Pi'=\exp\left[i\theta\left(\sigma_{+}\sigma_{-}+I_{z}+N/2\right)\right]$, indicating that it possesses $U\left(1\right)$ symmetry. The analytical solution of isotropic case has been discussed in detail in Ref.~\cite{PhysRevB.99.174308,PhysRevA.107.013714}. First of all, we calculate the squeezing parameters for the ground state in the isotropic case. Ref.~\cite{PhysRevA.107.013714} shows that there exists a a normal-to-superradiant phase transition in the limit of $\eta\rightarrow\infty$ and $N\rightarrow\infty$, and the critical point is $\tilde{g}_{c}=2/(1+\lambda)=2$. A detailed discussion is also provided in ~Ref.~\cite{PhysRevA.107.013714}. 
	
	For further analysis, we introduce the Dicke state $\left|n\right\rangle\equiv\left|\frac{N}{2},-\frac{N}{2}+n\right\rangle $ ($n\in\left[0,N\right]$), which is the eigenstate of the operators $\boldsymbol{I}^2$ and $I_{z}$. For $\eta\gg1$ and $N\gg 1$, the ground state is $\left|\downarrow,0\right\rangle \equiv\left|\downarrow\right\rangle \otimes\left|\frac{N}{2},-\frac{N}{2}\right\rangle $ when $\tilde{g}<2$, where $\left|\downarrow\right\rangle $ ($\left|\uparrow\right\rangle $) is the eigenstate of the operator $\sigma_{z}$. Immediately, we obtain $\xi^{2}_{S}=\xi^{2}_{R}=1$ under the condition of  $\tilde{g}<2$. For $\tilde{g}>2$, the ground state is given by~\cite{PhysRevA.107.013714}
	\begin{equation}\label{S2-11}
		\left|\psi_{-}\left(n\right)\right\rangle =\widetilde{P}_{\uparrow,n-1}^{-}\left|\uparrow,n-1\right\rangle +\widetilde{P}_{\downarrow,n}^{-}\left|\downarrow,n\right\rangle,
	\end{equation}
	where 
	\begin{align}
	\widetilde{P}_{\uparrow,n-1}^{-}&=\frac{\widetilde{\Omega}-\sqrt{1+\widetilde{\Omega}^{2}}}{\sqrt{2\left(1+\widetilde{\Omega}^{2}\right)-2\widetilde{\Omega}\sqrt{1+\widetilde{\Omega}^{2}}}},\label{S2-12}
	\\\widetilde{P}_{\downarrow,n}^{-}&=\frac{1}{\sqrt{2\left(1+\widetilde{\Omega}^{2}\right)-2\widetilde{\Omega}\sqrt{1+\widetilde{\Omega}^{2}}}},\label{S2-13}\\ 
	\widetilde{\Omega}&=\frac{\Omega-\omega}{\tilde{g}\sqrt{\left(N-n+1\right)n\Omega\omega}},\label{S2-14}
	\end{align}
    and 
    \begin{equation}\label{S2-15}
    	n_=\frac{\eta}{4}\left(\tilde{g}^2-\tilde{g}^{-2}\right).
    \end{equation}
	Here $n$ in Eq.~\eqref{S2-15} can be regarded as  the excitation
	number of the ground state. Utilizing the above equations, we can obtain the expression of spin squeezing parameters of the ground state for $\eta\gg1$, which is 
	\begin{equation}\label{S2-16}
		\xi_{S}^{2}=-\frac{2\left(n-\frac{N}{2}\right)^{2}}{N}+\frac{N}{2}+1,
	\end{equation}
	and 
	\begin{equation}\label{S2-17}
		\xi_{R}^{2}=-\frac{N}{2}+\frac{N^{2}\left(N+2\right)}{8\left(n-\frac{N}{2}\right)^{2}}.
	\end{equation}
	\begin{figure}[tbp]
		\includegraphics[width=\linewidth]{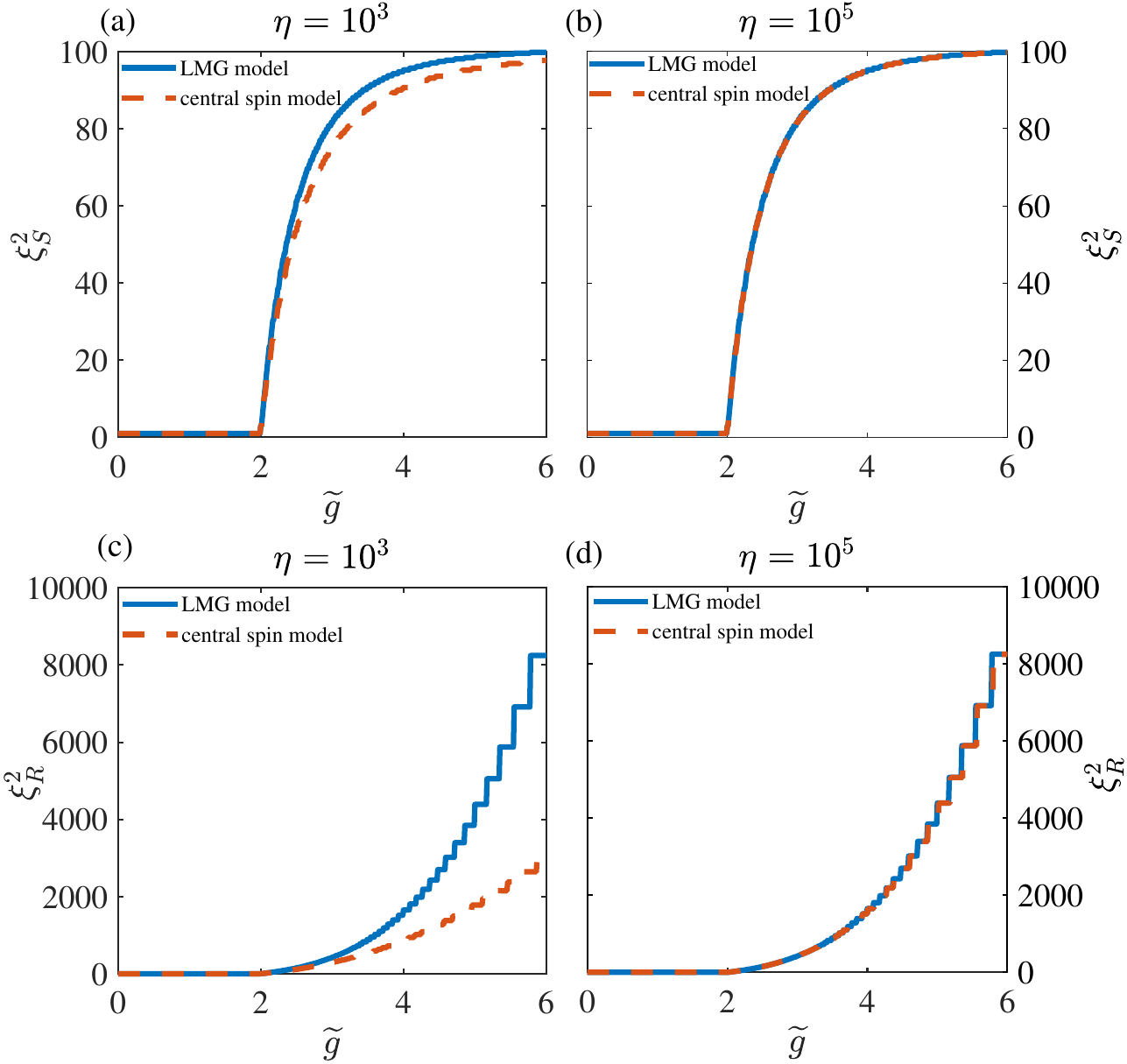}
		\caption{\label{fig:1} Spin squeezing parameters $\xi_{S}^2$ and $\xi_{R}^2$ of the ground state for the isotropic case as functions of $\tilde{g}$ with $N=200$. The first row [panels~(a)-(b)] corresponds  to the cases of spin squeezing parameter $\xi_{S}^{2}$ with $\eta=10^{3}$ and $\eta=10^{5}$, respectively. The second row [panels~(c)-(d)] corresponds  to the cases of spin squeezing parameter $\xi_{R}^{2}$ with $\eta=10^{3}$ and $\eta=10^{5}$, respectively. The solid line are the results of the LMG model and the dashed line are the results of the central spin model. 
		}
	\end{figure}
	The derivation of the above equations is provided in Appendix~\ref{Appendix:B}. The analytical results of Eq.~\eqref{S2-16} and Eq.~\eqref{S2-17} are shown in Fig.~\ref{fig:1}. One can see that the squeezing parameters $\xi_{S}^2$ and $\xi_{R}^2$  remain $1$ when $\tilde{g}<\tilde{g}_{c}$, and they begin to continuously increase after crossing the critical point $\tilde{g}_{c}$, which implies that the squeezing parameters can be used as an indicator to characterize the phase transition~\cite{PhysRevA.80.012318,MA201189}. Apart from this, we can also find that the squeezing parameters of the central spin model tend to approach those of the LMG model as the frequency ratio $\eta$ increases. As shown in Fig.~\ref{fig:1}(b) and Fig.~\ref{fig:1}(d), the two spin squeezing parameters are essentially consistent when the frequency ratio is $\eta=10^{5}$. 
	
	However, from the above analysis, we can find that the ground state of the central spin is not a spin-squeezed state due to  $\xi_{R}^2\geq\xi_{S}^2\geq1$. Subsequently, we will explore the possibility of generating spin-squeezed states through a dynamical approach. We choose $\left|\downarrow\right\rangle \otimes\left|\psi_{{\rm cs}}\right\rangle $ as the initial state, where $\left|\psi_{{\rm cs}}\right\rangle $ is the spin coherent state, i.e., $\left|\psi_{{\rm cs}}\right\rangle =\otimes_{k=0}^{N}\left[\cos\left(\theta_{0}/2\right)\left|\uparrow\right\rangle _{k}+e^{i\phi_{0}}\left(\theta_{0}/2\right)\left|\downarrow\right\rangle _{k}\right]$, and the exact solution of this dynamical process has been provided in \cite{PhysRevB.99.174308}, which is 
	\begin{align}\label{S2-18}
	\left|\psi_{f}\left(t\right)\right\rangle =&\sum_{m=0}^{N}\sqrt{C_{N}^{m}} e^{-i\left(-\frac{N}{2}+m-\frac{1}{2}\right)\omega t}\left(\cos\frac{\theta_{0}}{2}\right)^{m}\left(\sin\frac{\theta_{0}}{2}\right)^{N-m}\nonumber\\
	 & \times\left(P_{\downarrow}^{m}\left|\downarrow,m\right\rangle +P_{\uparrow}^{m}\left|\uparrow,m-1\right\rangle \right),
	\end{align}  
	where $P_{\downarrow}^{m}=i\Omega\sin\left(\Omega_{m}t/2\right)/\Omega_{m}+\cos\left(\Omega_{m}t/2\right)$,  $P_{\uparrow}^{m}=-i2\sqrt{k_{m}}A\sin\left(\Omega_{m}t/2\right)/\Omega_{m}$, and $\Omega_{m}=\sqrt{\Omega^{2}+4k_{m}A^{2}}$ with $k_{m}=m\left(N-m+1\right)$. Note that we set  $\phi_{0}=0$ for simplicity, and the mean spin direction is $\vec{n}_{0}=\left(\sin\theta\cos\phi,\sin\theta\sin\phi,\cos\theta\right)$, while the directions perpendicular to it are given by $\vec{n}_{1}=\left(-\sin\phi,\cos\phi,0\right)$ and $\vec{n}_{2}=\left(\cos\theta\cos\phi,\cos\theta\sin\phi,-\sin\theta\right)$. The spin squeezing parameter is given by
	\begin{equation}\label{S2-19}
		\xi_{S}^2=\frac{2\left(\mathcal{C}-\sqrt{\mathcal{A}^2+\mathcal{B}^2}\right)}{N},
	\end{equation}
	where $\mathcal{A}=\left\langle J_{\vec{n}_{1}}-J_{\vec{n}_{2}}\right\rangle$,
	$\mathcal{B}=\left\langle J_{\vec{n}_{1}}J_{\vec{n}_{2}}-J_{\vec{n}_{2}}J_{\vec{n}_{1}}\right\rangle $,
	and  $C=\left\langle J_{\vec{n}_{1}}^{2}+J_{\vec{n}_{2}}^{2}\right\rangle$. The values of the above parameters only depend on the following five parameters: $\left\langle I_{z}\right\rangle$, $\left\langle I_{z}^2\right\rangle$, $\left\langle I_{+}\right\rangle$, $\left\langle I_{+}^{2}\right\rangle$, $\left\langle I_{+}\left(2I_{z}+1\right)\right\rangle$~\cite{Jin_2009}. The specific expressions for the above parameters are presented in Appendix~\ref{Appendix:B}.  
	\begin{figure}[tbp] 
		\includegraphics[width=1.01\linewidth]{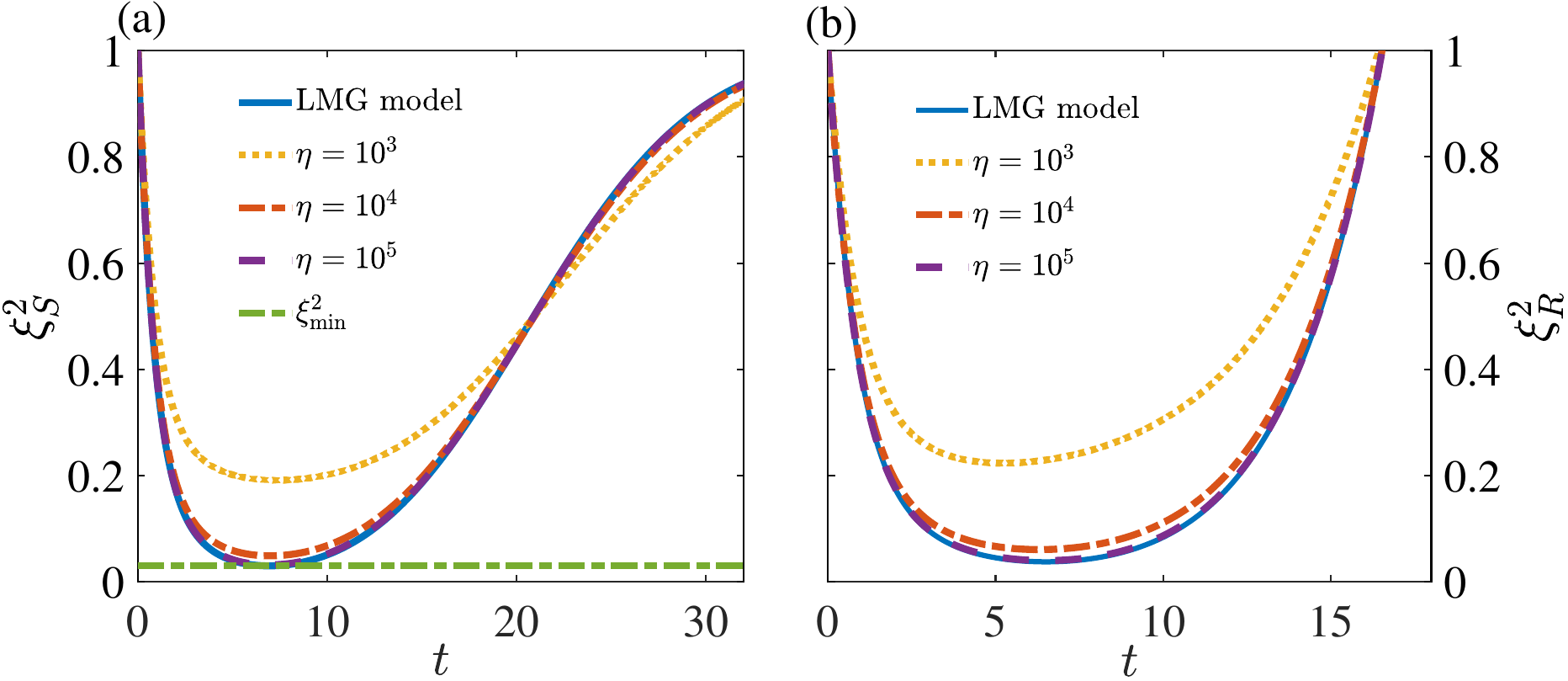}
		\caption{\label{fig:2} Time evolution of the spin squeezing parameters $\xi_{S}^2$ [panel~(a)] and $\xi_{R}^2$ [panel~(b)] for different frequency ratios $\eta$. The parameters chosen here are $N=200$, $\omega=1$ and $\tilde{g}=2$. The curves from top to bottom correspond to the spin squeezing parameter $\xi_{S}^{2}$ of the LMG model (blue solid line) and the central spin model with frequency ratios $\eta=10^{3}$ (yellow dotted line), $\eta=10^{4}$ (red dash-dot line), and $\eta=10^{5}$ (purple dashed line), respectively. In (a), the bottom straight line corresponds to the optimal spin squeezing parameter $\xi_{S,\rm min}$ (green dash-dot line).
		}
	\end{figure}
	
	In Fig.~\ref{fig:2}, we show the time evolution of spin squeezing parameters $\xi_{S}^2$ and $\xi_{R}^2$ of the central spin model with different frequency ratios $\eta$. Here we choose $\theta_{0}=\pi/2$ and $N$ is large enough ($N=200$),  and the optimal squeezing parameter for the above model is given by
	 \begin{equation}\label{S2-20}
	 \xi_{S,\rm min}^2\simeq\frac{1}{2}\left(\frac{N}{3}\right)^{-\frac{2}{3}},
	 \end{equation}
	 and the corresponding squeezing time is 
	 \begin{equation}\label{S2-21}
	 	t_{\rm min}\simeq\frac{4\cdot3^{\frac{1}{6}}N^{\frac{1}{3}}}{\tilde{g}^{2}\omega}.
	 \end{equation}
	 In fact, for the OAT interaction, i.e., $\chi I_{z}^2$, it has the following power-law scalings: $\xi_{S,\rm min}^2\sim N^{-2/3}$ and $\chi t_{\rm min}\sim N^{-2/3}$~\cite{PhysRevA.47.5138,PhysRevLett.102.100401,PhysRevLett.102.100401,Jin_2009,Huang2021}. From Eq.~\eqref{S2-20} and Eq.~\eqref{S2-21}, we can see that the power-law scalings of $\xi_{S,\rm min}^2$ and $t_{\rm min}$ (here $\chi=\tilde{g}^2\omega/4N$) are the same as the one-axis twisting case when $\eta$ is sufficiently large. On the other hand, a substantial detuning between the central spin and bath spins ($\Omega/\omega\gg1$) can induce an intra-species interaction, which is reported in Ref.~\cite{PhysRevLett.107.206806,PhysRevA.107.042613}. 
	 
	\subsection{\label{subsec:levelB}Anisotropic case} 
	
	Now we consider the anisotropic case, and without loss of generality we set $0<\lambda\leq1$. For the Hamiltonian in Eq.~\eqref{S2-2}, it is difficult to obtain its analytical solution. To further analyze it, we begin with the Hamiltonian in Eq.~\eqref{S2-7} ($\eta\rightarrow\infty$) and employ the mean-field approximation~\cite{PhysRevB.71.224420,PhysRevA.80.012318} to obtain the following Hamiltonian to obtain
	\begin{align}\label{S2-B1}
		H_{\rm MF}&=\left\langle \psi_{{\rm cs}}\left|H\right|\psi_{{\rm cs}}\right\rangle \nonumber\\ &=\frac{N\omega}{2}\cos\theta_{0}-\frac{\tilde{g}^{2}\omega\sin^{2}\theta_{0}}{4}\left(1+\lambda^{2}+2\lambda\cos2\phi_{0}\right),
	\end{align}
	where $\left|\psi_{{\rm cs}}\right\rangle $ is the spin coherent state. In order to minimize the energy, we find the following conditions: (\romannumeral1) for $\tilde{g}<2/\left(1+\lambda\right)$, we have $\theta_{0}=\pi$ and $\phi_{0}$ is arbitrary; (\romannumeral2) for $\tilde{g}>2/\left(1+\lambda\right)$, we have $\theta_{0}=\arccos\left(-\omega/\gamma_{x}\right)$, and $\phi_{0}=0, \pi$. Furthermore, we rotate the  
	the spin operators around the $y$ axis to align the $z$ axis with the direction of the semiclassical magnetization.  The rotated operators are expressed as $\tilde{I}_{x}=\cos\theta_{0}I_{x}-\sin\theta_{0}I_{z}$, $\tilde{I}_{y}=I_{y}$, and $\tilde{I}_{z}=\sin\theta_{0} I_{x}+\cos\theta_{0}I_{z}$, and the Hamiltonian becomes 
	\begin{align}\label{S2-B2}
		H_{\downarrow}=&-\frac{\gamma_{x}}{N}\left[\cos^{2}\theta_{0}\tilde{I}_{x}^{2}+\sin^{2}\theta_{0}\tilde{I}_{z}^{2}+\sin\theta_{0}\cos\theta_{0}\left(\tilde{I}_{x}\tilde{I}_{z}+\tilde{I}_{z}\tilde{I}_{x}\right)\right]\nonumber\\
		&-\frac{\gamma_{y}}{N}I_{y}^{2}+\gamma_{z}\left(-\sin\theta_{0}\tilde{I}_{x}+\cos\theta_{0}\tilde{I}_{z}\right).
	\end{align}
		\begin{figure}[tbp]
		\includegraphics[width=1.\linewidth]{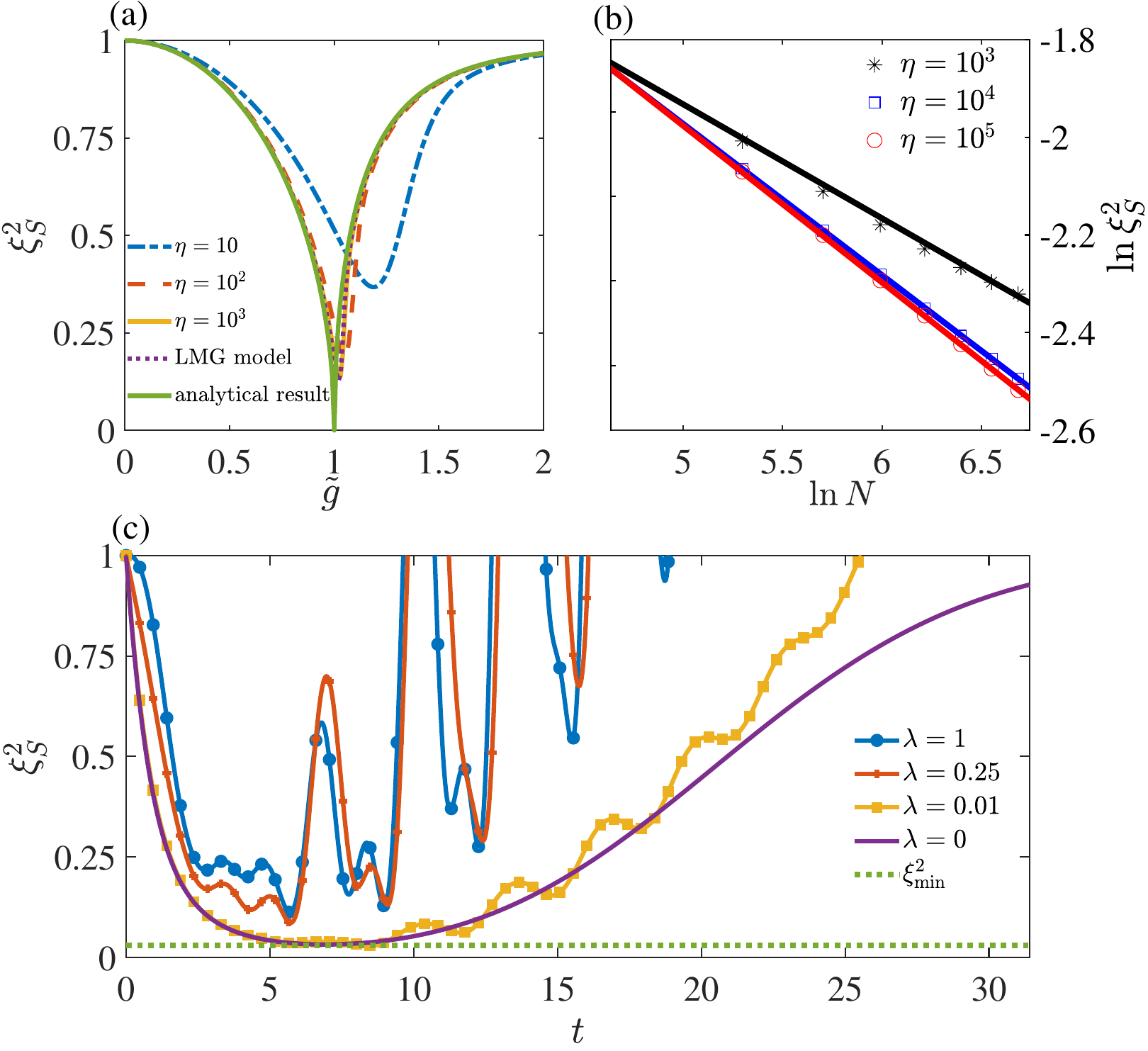}
		\caption{\label{fig:3} (a) Spin squeezing parameter $\xi_{S}^{2}$ of the ground state for the anisotropic case 
	 	as functions of $\tilde{g}$ with different frequency ratios $\eta$. Here we choose $N=200$, $\omega=1$ and $\lambda=1$. The curves from top to bottom correspond to the spin squeezing parameter $\xi_{S}^{2}$ of the central spin model with frequency ratios $\eta=10$ (blue dash-dot line), $\eta=10^{2}$ (red dashed line), and $\eta=10^{3}$ (yellow solid line), the LMG model (purple dotted line), and the analytical results given by Eq.~\eqref{S2-B7} and Eq.~\eqref{S2-B9} (solid green line), respectively. (b) Optimal squeezing parameters $\xi_{S,\rm min}^2$ of the ground state  as functions of $\ln N$ with different frequency ratios $\eta$. (c) Time evolution of spin squeezing parameters for different anisotropy parameters. The curves from top to bottom correspond to $\lambda=1$ (blue circle line), $\lambda=0.25$ (blue circle line), $\lambda=0.01$ (yellow point line), $\lambda=0$ (purple solid line). 
		}
	\end{figure}
	Then we apply the Holstein-Primakoff transformation to the above equation, i.e.,
	$\tilde{I}_{+}=\sqrt{N}a$, $\tilde{I_{-}}=\sqrt{N}a^{\dagger}$, and $\tilde{I}_{z}=N/2-a^{\dagger}a$, and  assume that $N$ is sufficiently large ($\sqrt{a^{\dagger}a/N}\simeq1$), then Eq.~\eqref{S2-B2} can be rewritten as 
	\begin{equation}\label{S2-B3}
	H_{\downarrow}=-\frac{\gamma_{x}}{2}\cos^{2}\theta_{0}x^{2}-\frac{\gamma_{y}}{2}p^{2}+\left(\gamma_{x}\sin^{2}\theta_{0}-\omega\cos\theta_{0}\right)a^{\dagger}a,
	\end{equation}
	where $x=\left(a^{\dagger}+a\right)/\sqrt{2}$ and $p=i\left(a^{\dagger}-a\right)/\sqrt{2}$. In the above derivation, we utilize $\omega\sin\theta_{0}+\gamma_{x}\sin\theta_{0}\cos\theta_{0}=0$ and neglect the constant term. 
	
	Similarly, the spin squeezing parameter differs before and after the critical point. 
	First of all, for $\tilde{g}<2/\left(1+\lambda\right)$ and $\theta_{\theta}=\pi$, Eq.~\eqref{S2-B3} becomes 
	\begin{equation}\label{S2-B4}
     H=\omega a^{\dagger}a-\frac{1}{2}\left(\gamma_{x}x^{2}+\gamma_{y}p^{2}\right).		
	\end{equation}
	We use a unitary transformation with a squeezed operator $\mathcal{S}\left(r\right)=\exp\left[\left(r/2\right)\left(a^{2}-a^{\dagger2}\right)\right]$ to Eq.~\eqref{S2-B3} and obtain
	\begin{equation}\label{S2-B5}
		H=\sqrt{\left(\omega-\gamma_{x}\right)\left(\omega-\gamma_{y}\right)}\left(a^{\dagger}a+\frac{1}{2}\right),
	\end{equation}
	where $r=\frac{1}{4}\ln\left[\left(\omega-\gamma_{x}\right)/\left(\omega-\gamma_{y}\right)\right]$. It is worth noting that the ground state of the Hamiltonian in Eq.~\eqref{S2-7} satisfies that  $\left\langle \left\{ I_{x},I_{y}\right\} \right\rangle=0$, thus spin squeezing parameter is given by~\cite{PhysRevA.80.012318} 
	\begin{equation}\label{S2-B6}
		\xi_{S}^{2}=\frac{4{\rm min}\left\{ \left\langle I_{x}^{2}\right\rangle ,\left\langle I_{y}^{2}\right\rangle \right\} }{N}.
	\end{equation}
	Utilizing the above equations, we ultimately obtain the spin squeezing parameter is given by
	\begin{equation}\label{S2-B7}
		\xi_{S}^{2}=\left[\frac{\tilde{g}_{c}^{2}-\tilde{g}^{2}}{\tilde{g}_{c}^{2}-\gamma\tilde{g}^{2}}\right]^{\frac{1}{2}},
	\end{equation}
	where $\gamma=\gamma_{y}/\gamma_{x}$.
	And for $\tilde{g}>2/\left(1+\lambda\right)$ and $\theta_{0}=\arccos\left(-\omega/\gamma_{x}\right)$, then we have  
	\begin{equation}\label{S2-B8}
		H=\omega\sqrt{\tilde{g}^{2}\lambda\left(\frac{\gamma_{x}}{\omega}-\frac{\omega}{\gamma_{x}}\right)}\left(a^{\dagger}a+\frac{1}{2}\right)-\frac{\gamma_{x}}{2},
	\end{equation}
	and $r=\frac{1}{4}\ln\left[\left(\gamma_{x}/\omega-\omega/\gamma_{x}\right)/\tilde{g}^{2}\lambda\right]$. The corresponding spin squeezing parameter is 
	\begin{equation}\label{S2-B9}
	\xi_{S}^{2}=\left[\frac{1}{\tilde{g}^{2}\lambda}\left(\frac{\tilde{g}^{2}}{\tilde{g}_{c}^{2}}-\frac{\tilde{g}_{c}^{2}}{\tilde{g}^{2}}\right)\right]^{\frac{1}{2}}.
	\end{equation}
	From Eq.~\eqref{S2-B7} and Eq.~\eqref{S2-B9}, we can see that $\xi_{S}^2<1$ when $\tilde{g}<2/\sqrt{\left(1+\lambda\right)\left(1-\lambda\right)}$, which implies that one can generate the spin-squeezed state by preparing the ground state of the anisotropic central spin model with a large frequency ratio $\eta$. In Fig.~\ref{fig:3}~(a), we present the spin squeezing parameter $\xi_{S}^2$ of  the ground state for the anisotropic case 
	varies with $\tilde{g}$. It can be seen that  the spin squeezing parameter $\xi_{S}^2$ of the anisotropic central spin model tends to that of the LMG model as the frequency ratio $\eta$ increases.
	
	In Fig.~\ref{fig:3}~(a), it is evident that the optimal squeezing parameter appears in the vicinity of the critical point. However, one can see that the analytical expression of $\xi_{S}^{2}$ (solid green line) decreases to zero at the critical point, which means the results of Eq.~\eqref{S2-B7} and Eq.~\eqref{S2-B9} fail when $\tilde{g}$ approach $\tilde{g}_{c}$. In such a case, it is necessary to consider the higher-order corrections.  We utilize numerical simulations to calculate the variation of the optimal spin squeezing parameter $\xi_{S,\rm min}^2$ with respect to $N$ and find that $\xi_{S,\rm min}^2\sim N^{-1/3}$ in the limit of $\eta\rightarrow\infty$, which is consistent with the result of the LMG model in Ref.~\cite{PhysRevA.69.022107,MA201189}.
	 
	 The spin squeezing parameters of the dynamic process with various $\lambda$ are also investigated by numerical simulation. In Fig.\ref{fig:3}~(c), we compare the dynamical evolution of spin squeezing parameter $\xi_{S}^2$ with different values of $\lambda$. It is clear that the smaller the value of $\lambda$, the smaller the corresponding optimal squeezing parameter. In other words, for $\lambda=0$, the associated optimal squeezing parameter $\xi_{S,\rm min}^2$ is minimal.
	 
	 In summary, our analysis of the above results reveals that, under the condition of fixed $N$, the most effective method for generating the optimal spin-squeezed state is through the dynamic evolution of the isotropic central spin model. In this process, it is crucial to make the ratio $\eta$ between the frequency of the central spin and that of the bath spin  as large as possible.
	
	 \begin{figure}[tbp]
	 	\includegraphics[width=1.05\linewidth]{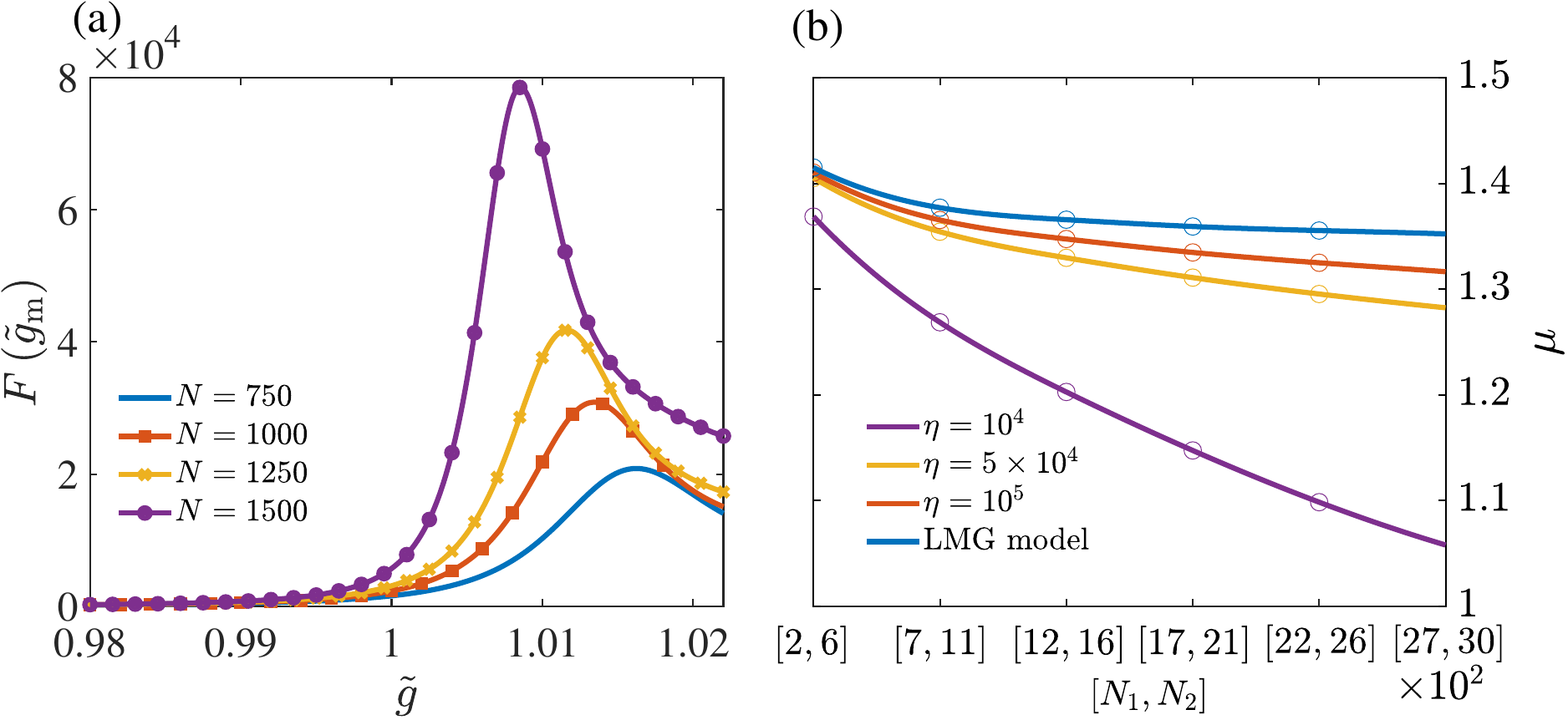}
	 	\caption{\label{fig:4} (a) QFI  of the ground state 
	 		as functions of $\tilde{g}$ with different numbers of bath spins $N$. Here we choose $\eta=10^{5}$, $\omega=1$, $\lambda=1$ and $\tilde{g}_{c}=1$. The curves from top to bottom correspond to the QFI of the central spin model with  $N=750$ (purple line), $N=1000$ (yellow line), and $N=1250$ (red line) and $N=1250$ (blue line), respectively. (b) critical exponent $\mu$ within various system size $\left[N_{1},N_{2}\right]$. The curves from top to bottom correspond to the cases of $\eta=1000$ (blue line), $\eta=10000$ (red line), and $\eta=50000$ (yellow line), respectively. In addition, the case of the LMG model is also presented  for comparison (top blue line). 
	 	}
	 \end{figure}
	 \section{\label{sec:levelIII} QUANTUM FISHER INFORMATION AND FINITE-SIZE ANALYSIS }
	 Quantum Fisher information is a core concept in quantum metrology, and the QFI of the ground state can be regard as an indicator to detect quantum phase transition even without knowing the information related to  order parameter~\cite{doi:10.1142/S0217979210056335}. Therefore, in this section, we will utilize QFI to analyze the finite-size behavior of the anisotropic central spin model. The finite-size analysis of the isotropic case ($\lambda=0$) has been discussed in Ref.~\cite{PhysRevA.107.013714}, thus here we will focus on the anisotropic case ($\lambda\neq0$).  
	 
	 For a pure state $\left|\psi\left(\tilde{g}\right)\right\rangle$, the QFI with respect to parameter $\tilde{g}$ is give by
	 \begin{equation}\label{S4-1}
	 	F\left(\tilde{g}\right)=4\left\langle \partial_{\tilde{g}}\psi|\partial_{\tilde{g}}\psi\right\rangle -4\left|\left\langle \partial_{\tilde{g}}\psi|\psi\right\rangle \right|^{2},
	 \end{equation}                                  
	 and its value is four times the  fidelity susceptibility~\cite{doi:10.1142/S0217979210056335,PhysRevA.102.063721}. 
	 According to the conclusion in Ref.~\cite{PhysRevE.78.032103}, for the LMG model, the QFI of the ground state around the critical point presents scaling behavior as 
	 \begin{equation}\label{S4-2}
	 	F\left(\tilde{g}_{\rm m}\right)\propto N^{\mu},
	 \end{equation}
	 where $\tilde{g}_{\rm m}$ is the position of the maximal QFI, $N$ is the size of the system, and $\mu$ is the critical exponent with a specific value of $\mu=4/3$.  For the anisotropic central spin model, two conditions are required for a phase transition to occur, namely, the frequency ratio $\eta$ and the size of the system $N$ tend to infinity. Nevertheless, when the frequency ratio $\eta$ is finite, the scaling behavior of  $F\left(\tilde{g}_{\rm m}\right)$ in Eq.~\eqref{S4-2} is still unknown. Therefore, we utilize  the exact diagonalization to obtain the ground state and calculate the critical exponent $\mu$ for a finite $\eta$. 
	 
	 In Fig.~\ref{fig:4}~(a), we depict the variation of  the QFI of the ground state for the anisotropic central spin model  with different numbers of bath spins. It is observed that as the $N$ increases, the maximum values of the QFI exhibits a progressively growing trend, and their corresponding positions gradually approach the critical point. The critical exponent $\mu$ within different system size ranges $[N_{1}, N_{2}]$ is provided in Fig.~\ref{fig:4}~(b). It is evident that, as the frequency ratio $\eta$ increases, the critical exponent converges to $4/3$ when $N$ is sufficiently large, which corresponds precisely to the critical exponent of the LMG model (solid bule line)~\cite{PhysRevE.78.032103}. On the other hand, we find that if $\eta$ is not sufficiently large, the critical exponent will rapidly decrease with increasing $N$ until it converges to zero.  Namely, if $N$ tends to infinity while $\eta$ remains finite, the corresponding critical exponent $\mu$ will decay to zero, which implies that  the quantum phase transition will not occur. Hence, the condition for the phase transition to occur requires  $\eta$ and $N$ both tend towards infinity.
	 
	 \section{\label{sec:levelIIII} CONCLUSION }
	 In summary, we have analyzed the spin squeezing and quantum phase transition in the anisotropic central spin model. Utilizing the Schrieffer-Wolff transformation, we analytically  established the mapping  between the anisotropic central spin model and the anisotropic  LMG  model.  Meanwhile, it is shown that the substantial detuning between the central spin and bath spins  can induce an intra-species nonlinear interaction. Inspired by the above results, we investigated the potential pathways for generating spin-squeezed states with central spin systems. 
Through comparisons, we found that the dynamics initiated with a spin coherent state  can generate the optimal spin-squeezed state. Furthermore, when the anisotropy parameter $\lambda$ is zero, the optimal spin squeezing parameter scales with the system size as $N^{-2/3}$. The critical exponent $\mu$ of the quantum Fisher information around the critical point has also been employed to study the finite-size behavior in the anisotropic central spin model. Specifically, we obtained  $\mu=4/3$ as $\eta$ and $N$ approach infinity, however, when $\eta$ is finite, it is found that the value of $\mu$ decreases to zero with the increase of $N$. This work provides a new potential pathway for generating spin-squeezed states in practical physical systems, simultaneously offering a fresh perspective for our understanding of  the relationship between spin squeezing and quantum criticality.

	\begin{acknowledgments}
		We acknowledge Dr.~Zhucheng Zhang for valuable suggestions. This work was supported by the National Natural Science Foundation of China (Grants No. 12088101 and No. U2330401) and Science Challenge Project (Grant No.
		TZ2018005).
	\end{acknowledgments}
	\appendix
	
	\section{DERIVATION OF MAPPING  BEWTEEN CENTRAL SPIN MODEL AND LMG MODEL}\label{Appendix:A}
	In this section, we give the derivation of  the mapping between the central spin model and the LMG model. First of all,  Eq.~\eqref{S2-2} can be rewritten as 
	\begin{equation}\label{A1}
	H=H_{0}+AV,
	\end{equation}  
	where 
	\begin{align}
	H_{0}=&\frac{\Omega}{2}\sigma_{z}+\omega I_{z},\label{A2}\\
	V=&\left(1+\lambda\right)J_{x}\sigma_{x}+\left(1-\lambda\right)J_{y}\sigma_{y}.	\label{A3}
	\end{align}
	Note that $H_{0}$ is diagonal with respect to the spin subspace while  $V$ is off-diagonal. Then we perform the Schrieffer-Wolff transformation $e^{S}$ on Eq.~\eqref{A1} and obtain~\cite{PhysRevLett.115.180404,PhysRevLett.117.123602}
	\begin{equation}\label{A4}
	H’=e^{-S}He^{S}=\sum_{k=0}^{\infty}\frac{1}{k!}\left[H,S\right]^{\left(k\right)},
	\end{equation}
	where  $\big[H,S\big]^{\left(k\right)}=\big[\big[H,S\big]^{\left(k-1\right)},S\big]$, $\big[H,S\big]^{\left(0\right)}=H$, and the generator $S$ is an anti-Hermitian and block-off-diagonal operator. In order to diagonalize the Hamiltonian $H'$, we need the block-off-diagonal part of $H'$ to be zero up to second order in $A$, which leads to the follow relationship
	\begin{equation}\label{A5}
	\left[H_{0},S\right]=-V,
	\end{equation}
	and we find the expression of $S$ satisfying Eq.~\eqref{A5} is as follows
	\begin{equation}
		S=-i\frac{\left(1+\lambda\right)}{\Omega}I_{x}\sigma_{y}+i\frac{\left(1-\lambda\right)}{\Omega}I_{y}\sigma_{x}.
	\end{equation}
	In the limit of $\eta\rightarrow\infty$, the  Hamiltonian in Eq.~\eqref{A4} becomes 
	\begin{align}
		H'=&H_{0}+\frac{A^{2}}{2}\left[V,S\right]\nonumber\\
		  =&\frac{\Omega}{2}\sigma_{z}+\left[\omega-A^{2}\frac{\left(1+\lambda\right)\left(1-\lambda\right)}{\Omega}\right]I_{z}\nonumber\\
		  &+A^{2}\left[\frac{\left(1+\lambda\right)^{2}}{\Omega}I_{x}^{2}\sigma_{z}+\frac{\left(1-\lambda\right)^{2}}{\Omega}I_{y}^{2}\sigma_{z}\right].
	\end{align}
	Furthermore, we set $\tilde{g}=2A\sqrt{N/\Omega\omega}$, $\gamma_{x}=\tilde{g}^{2}\omega\left(1+\lambda\right)^{2}/4$, $\gamma_{y}=\tilde{g}^{2}\omega\left(1-\lambda\right)^{2}/4$, and finally we get the Hamiltonian in Eq.~\eqref{S2-3}, which precisely corresponds to the Hamiltonian of the anisotropic LMG model.
	\section{DERIVATION OF SPIN SQUEEZING PARAMETER IN DYNAMICAL PROCESS }\label{Appendix:B}
	First of all, we give the derivation of Eq.~\eqref{S2-16} and
	Eq.~\eqref{S2-17}. For the ground state of the LMG model, the mean spin direction is always along the $z$-axis, thus we have $\left\langle I_{x}\right\rangle =\left\langle I_{y}\right\rangle =0$, and then we obtain~\cite{MA201189}
	\begin{align}\label{B1}
		{\rm min}\left(\Delta^{2}I_{\vec{n}_{\bot}}\right)&=
		\frac{1}{2}\left[\left\langle I_{x}^{2}+I_{y}^{2}\right\rangle -\sqrt{\left(\left\langle I_{x}^{2}-I_{y}^{2}\right\rangle \right)^{2}+4{\rm Cov}\left(I_{x},I_{y}\right)^{2}}\right]\nonumber\\
		&=\frac{1}{2}\left\langle I_{x}^{2}+I_{y}^{2}\right\rangle, 
	\end{align} 
	where ${\rm Cov}\left(I_{x},I_{y}\right)=\frac{1}{2}\left\langle I_{x}I_{y}+I_{y}I_{x}\right\rangle $. 
	
	For $\eta\rightarrow\infty$, we make an approximation that $\widetilde{P}_{\uparrow,n-1}^{-}\simeq0,\widetilde{P}_{\downarrow,n}^{-}\simeq1$, and Eq.~\eqref{B1} becomes 
	\begin{align}\label{B2}
		{\rm min}\left(\Delta^{2}I_{\vec{n}_{\bot}}\right)&=
		\frac{1}{2}\left\langle \boldsymbol{I}^{2}-I_{z}^{2}\right\rangle \nonumber
		\\&=-\frac{1}{2}\left(n-\frac{N}{2}\right)^{2}+\frac{N^{2}}{8}+\frac{N}{4}.
	\end{align}	 
	Utilizing Eq.~\eqref{B2} we can get Eq.~\eqref{S2-16} and
	Eq.~\eqref{S2-17}.

	 Then we give the derivation of the time evolution of spin squeezing parameters $\xi_{S}^2$ and $\xi_{R}^2$. The initial state we choose is given in Eq.~\eqref{S2-18}, and under the condition of $\eta\gg1$ we have 
	\begin{equation}\label{B3}
	\left\langle I_{z}\right\rangle=\frac{N}{2}\cos\theta_{0},	
	\end{equation}
	\begin{equation}\label{B4}
			\left\langle I_{z}^{2}\right\rangle =\frac{1}{8}N\left[N+1+\left(N-1\right)\cos2\theta_{0}\right].	
	\end{equation}

	 Then we make the following approximations: $\Omega_{n}\simeq\Omega$, $P_{\downarrow}^{n}\simeq e^{i\frac{\Omega_{n}t}{2}}$, $P_{\uparrow}^{n}\simeq0$ and $\Omega_{n}\simeq\Omega+2k_{n}A^{2}$. And we obtain
	 \begin{widetext}
	 \begin{align}
	 	\left\langle I_{+}\right\rangle &\simeq\sum_{n=0}^{N-1}C_{N-1}^{n}N\left(\cos^{2}\frac{\theta_{0}}{2}\right)^{n}\left(\sin^{2}\frac{\theta_{0}}{2}\right)^{N-1-n}e^{-i\frac{\left(N-2n\right)A^{2}t}{\Omega}}e^{i\omega t}\cot\frac{\theta_{0}}{2}\nonumber\\
	 	&=\frac{1}{2}Ne^{i\left(\omega-\frac{A^{2}}{\Omega}\right)t}\left(\cos\frac{A^{2}t}{\Omega}+i\sin\frac{A^{2}t}{\Omega}\cos\theta_{0}\right)^{N-1}\sin\theta_{0},\label{B5}
	 \end{align}
	 \begin{align}
	 \left\langle I_{+}^{2}\right\rangle &\simeq e^{2i\omega t}\sum_{n=0}^{N-1}\left(\cos^{2}\frac{\theta_{0}}{2}\right)^{n}\left(\sin^{2}\frac{\theta_{0}}{2}\right)^{N-n}C_{N-2}^{n}N\left(N-1\right)e^{\frac{-i2\left(N-2n-1\right)A^{2}t}{\Omega}}\cot^{2}\frac{\theta_{0}}{2}\nonumber\\
	 &=\frac{1}{4}e^{2i\left(\omega-\frac{A^{2}}{\Omega}\right)t}N\left(N-1\right)\left[\cos\left(\frac{2A^{2}t}{\Omega}\right)+i\cos\theta_{0}\sin\left(\frac{2A^{2}t}{\Omega}\right)\right]^{N-2}\sin^{2}\theta_{0},\label{B6}
	\end{align}
	   \begin{align}
	  	2\left\langle I_{+}I_{z}\right\rangle +\left\langle I_{+}\right\rangle &\simeq\sum_{n=0}^{N-1}C_{N-1}^{n}N\left(\cos^{2}\frac{\theta_{0}}{2}\right)^{n}\left(\sin^{2}\frac{\theta_{0}}{2}\right)^{N-n}e^{-i\frac{\left(N-2n\right)A^{2}t}{\Omega}}e^{i\omega t}\left[2\left(-\frac{N}{2}+n\right)+1\right]\cot\frac{\theta_{0}}{2}\nonumber\\
	  	&=\frac{1}{2}e^{i\left(\omega-\frac{A^{2}}{\Omega}\right)t}N\left(N-1\right)\left[\cos\left(\frac{A^{2}t}{\Omega}\right)+i\cos\theta_{0}\sin\left(\frac{A^{2}t}{\Omega}\right)\right]^{N-2}\left[\cos\theta_{0}\cos\left(\frac{A^{2}t}{\Omega}\right)+i\sin\left(\frac{A^{2}t}{\Omega}\right)\right]\sin\theta_{0}.\label{B7}
	  \end{align}
	\end{widetext}	
	  For the sake of simplicity, we choose $\theta_{0}=\pi/2$, and the directions perpendicular to the mean spin direction
	  $\vec{n}_{0}$ are $\vec{n}_{1}=\left(-\sin\phi,\cos\phi,0\right)$, $\vec{n}_{2}=\left(0,0,1\right)$ and $\phi=\omega t$.
	  Therefore, the spin squeezing parameter is given by~\cite{MA201189}
	  \begin{align}
	  	\xi_{S}^2&=\frac{2}{N}\left[\left\langle I_{\vec{n}_{1}}^{2}+I_{\vec{n}_{2}}^{2}\right\rangle -\sqrt{\left(\left\langle I_{\vec{n}_{1}}^{2}-I_{\vec{n}_{2}}^{2}\right\rangle \right)^{2}+4{\rm Cov}\left(I_{\vec{n}_{1}},I_{\vec{n}_{2}}\right)^{2}}\right]\nonumber\\
	  	&=\frac{2}{N}\left(\mathcal{C}-\sqrt{\mathcal{A}^{2}+\mathcal{B}^{2}}\right),
	  \end{align}
	 where 
	  \begin{align}
	 	\mathcal{A}&=\frac{1}{8}\left(N-1\right)N\left[1-\cos\left(\frac{A^{2}t}{\Omega}\right)^{N-2}\right]\label{B9},\\
	 	\mathcal{B}&=\frac{N}{2}\left(N-1\right)\cos\left(\frac{A^{2}t}{\Omega}\right)^{N-2}\sin\left(\frac{A^{2}t}{\Omega}\right),\label{B10}\\
	 	\mathcal{C}&=\mathcal{A}+\frac{N}{2}.\label{B11}
	 \end{align}
	 By substituting Eq.~\eqref{B9}, Eq.~\eqref{B10}, and Eq.~\eqref{B11} into Eq.~\eqref{S2-19}, we can obtain the analytical solution for the time evolution of the spin squeezing parameter $\xi_{S}^2$.
	\bibliography{mybib}
\end{document}